\pdfoutput=1

\RequirePackage{fix-cm}
\documentclass[twocolumn,epjc3]{svjour3}
\RequirePackage[T1]{fontenc}
\RequirePackage{mathptmx}
\smartqed

\usepackage[english]{babel}
\usepackage{multirow}
\usepackage{makecell}
\usepackage{amsmath,amssymb, gensymb}
\usepackage[breaklinks=true]{hyperref}
\usepackage{url}
\usepackage{siunitx}
\usepackage{xspace}
\usepackage{microtype}
\usepackage{subfig}
\usepackage[useregional]{datetime2}
\usepackage{graphicx}
\usepackage{textcomp}
\usepackage{booktabs}
\usepackage{pgfplots}
\usetikzlibrary{patterns}
\pgfplotsset{compat=1.18}

\newcommand{\bbonu}{\ensuremath{\beta\beta0\nu}}
\newcommand{\Qbb}{\ensuremath{Q_{\beta\beta}}}
\newcommand{\stwo}{\ensuremath{S_2}}
\newcommand{\sone}{\ensuremath{S_1}}

\journalname{Eur. Phys. J. C}

\makeatletter
\let\makeheadbox\relax % remove the "manuscript No. ... editor" header box
\date{}                % remove the "the date of receipt ..." line
\makeatother

\begin{document}

\title{Evaluation of tracking plane pitch and xenon-helium gas mixtures for the NEXT experiment}

% Author list
\author{
 NEXT Collaboration\\\\
M.~P\'erez Maneiro\thanksref{addr21}
\and
M.~Cid\thanksref{addr15,addr21}
\and
J.D.~Villamil\thanksref{addr15}
\and
S.~Torelli\thanksref{addr17}
\and
N.~L\'opez-March\thanksref{addr15}
\and
J.~Renner\thanksref{f1,addr15}
\and
S.R.~Soleti\thanksref{addr17,addr7}
\and
J.A.~Hernando~Morata\thanksref{addr21}
\and
J.J.~G\'omez-Cadenas\thanksref{f2,f3,addr17,addr7}
\and
H.~Almaz\'an\thanksref{addr17}
% ...
}

\thankstext{f1}{email: jrenner@ific.uv.es (corresponding author)}
\thankstext{f2}{email: jjgomezcadenas@dipc.org (corresponding author)}
\thankstext{f3}{NEXT Spokesperson.}
\thankstext{f4}{Deceased.}
\thankstext{f5}{Now at Weizmann Institute of Science, Israel.}
\thankstext{f6}{Now at Los Alamos National Laboratory, USA.}

\institute{
% ...
Ikerbasque (Basque Foundation for Science), Bilbao, E-48009, Spain \label{addr7}
% ...
\and
Instituto de F\'isica Corpuscular (IFIC), CSIC \& Universitat de Val\`encia, Calle Catedr\'atico Jos\'e Beltr\'an, 2, Paterna, E-46980, Spain \label{addr15}
% ...
\and
Donostia International Physics Center, BERC Basque Excellence Research Centre, Manuel de Lardizabal 4, San Sebasti\'an / Donostia, E-20018, Spain \label{addr17}
% ...
\and
Instituto Gallego de F\'isica de Altas Energ\'ias, Univ.\ de Santiago de Compostela, Campus sur, R\'ua Xos\'e Mar\'ia Su\'arez N\'u\~nez, s/n, Santiago de Compostela, E-15782, Spain \label{addr21}
% ...
}

\maketitle

% Abstract
\begin{abstract}
NEXT is an experimental program to search for neutrinoless double-beta decay (\bbonu) in high-pressure xenon gas using electroluminescent time projection chambers. A key feature of the NEXT technology is the reconstruction of MeV-scale electron tracks, which plays a key role in identifying \bbonu\ events. This work presents a Monte Carlo study aimed at determining the optimal configuration for NEXT detectors employing SiPM-based tracking planes. The simulations are carried out in the NEXT-100 detector geometry, and to classify the events as signal or background, we use a Sparse Convolutional Neural Network. We find that topological rejection power improves by approximately a factor of two at the lower pressure studied (5 bar) compared to the baseline design (13.5 bar). A finer pitch also offers consistent improvement, where reducing pitch from 15 mm to 5 mm decreased the accepted background fraction by a factor of approximately 1.4 at 13.5 bar for a fixed signal efficiency in the different configurations studied. The addition of helium to reduce electron diffusion provides a comparable enhancement, reducing the accepted background fraction by up to a factor of 1.5 at 13.5 bar. We highlight two key practical configurations at 13.5 bar pressure: the present NEXT-100 (pure Xe with a 15 mm tracking plane pitch) and an upgraded NEXT-100 (a Xe-He gas mixture and a 10 mm tracking plane pitch). These results provide quantitative guidance for the design and optimisation of future NEXT detectors, balancing physics performance with technical and financial constraints.
\end{abstract}

\section{Introduction}\label{s:intro}

The observation of neutrinoless double-beta (\bbonu) decay is still one of the highest-priority goals in particle physics. Its discovery would prove that the neutrino is Majorana in nature (its own antiparticle), establish the violation of lepton number conservation, and provide some insight into the absolute neutrino mass scale~\cite{Avignone2008,Dolinski2019,Gomez-Cadenas:2023vca}. The search for this decay requires ultra-low background experiments with large isotopic masses and excellent energy resolution to distinguish a potential signal peak from background events at the Q-value of the decay (\Qbb).

The present generation of experiments has pushed the \bbonu\ sensitivity to half-lives beyond $10^{26}$~years, with the most stringent limit set by KamLAND-Zen at $T_{1/2} > 3.8 \times 10^{26}$ yr for $^{136}$Xe~\cite{KamLAND-Zen:2025}. Other leading efforts include the LEGEND experiment, which has set a limit of $T_{1/2} > 1.9 \times 10^{26}$ yr for $^{76}$Ge~\cite{LEGEND2025}.

The Neutrino Experiment with a Xenon TPC (NEXT) is a series of electroluminescent high-pressure xenon time projection chambers (HPXe-EL TPCs) designed to search for \bbonu\ in $^{136}\mathrm{Xe}$. The first radiopure apparatus of the NEXT series was NEXT-White, which deployed a mass of 5 kg enriched to 90\% $^{136}$Xe when operating at 10 bar pressure. The detector was in stable operation at the Laboratorio Subterráneo de Canfranc (LSC) from October 2016 to July 2021, demonstrating unambiguously the excellent performance of the HPXe technology for \bbonu\ searches~\cite{NEXT:2018zho,Novella:2019cne}, including: an energy resolution of $\sim1\%$ FWHM at 2.6 MeV~\cite{Renner:2019pfe,NEXT:2024btf}, an efficient discrimination between one- and two-electron tracks by means of the event topology~\cite{Ferrario:2019kwg,Kekic2021,NEXT:2021pjq}, a measurement of the $^{136}\mathrm{Xe}$ $\beta\beta2\nu$ half-life~\cite{NEXT:2021dqj}, and a demonstration of the potential of the technology for \bbonu~searches~\cite{NEXT:2023daz}. The larger NEXT-100 detector~\cite{NEXT:2025yqw}, currently in operation at the LSC, scales up the technology to hold $\sim$100 kg of enriched xenon and has already demonstrated $<$ 1\% energy resolution at \Qbb~\cite{NEXT_2025_energy,NEXT_2026_Kr} in a commissioning phase. The planned NEXT-HD apparatus~\cite{NEXT:2020amj} will further scale the mass to about several tons at 15 bar. 

Optimising the design of these future large-scale detectors requires a scalable method to evaluate their topological discrimination power. Double beta decay events, consisting of two energetic electrons emitted from a common vertex, produce a track topology that is notably different from that of one-electron events produced by gamma ray backgrounds. We choose to employ a Sparse Convolutional Neural Network (SCN)~\cite{Graham2018}, given the sparse nature of the energetic electron tracks in xenon gas, which can be trained on simulated datasets to separate signal (\bbonu) and background (one-electron) events. This follows previous work carried out within NEXT employing neural networks, in particular \cite{Kekic2021}, in which an SCN was trained on simulated events and applied to experimental data from the NEXT-White detector. Various other experiments have used the SCN, such as MicroBooNE \cite{MicroBoone_SCN_2021}, NOvA \cite{Shmakov_2023}, and DUNE simulation studies \cite{AbedAbud_2023}, for event classification and segmentation. Here we use the SCN to quantify the ability of the detector to distinguish between signal and background events of the same energy under various configurations of tracking plane pitch, pressure, and electron diffusion.

\subsection{Operating principles}
The principle of operation of an HPXe-EL TPC is based on several distinct properties of xenon: a) xenon scintillates (with an average wavelength of 172 nm, well into the VUV spectrum) due to the formation and decay of short-lived excimers, when a charged particle excites the gas. Therefore, the primary scintillation produced by the charged particle, denoted as \sone, is not re-absorbed by the gas and can be observed; b) xenon is a noble gas, and therefore xenon atoms have a closed electron shell, so drifting electrons are not captured by neutral atoms. Given sufficiently high gas purity, electrons can be drifted over long distances of up to meters by a moderate (several hundred V/cm) electric field. As a consequence, a track of ionisation electrons can be transported with minimal attachment losses towards the TPC anode, where it crosses a region of higher electric field (the EL region); c) in the EL process, electrons gain enough energy to excite xenon atoms but not enough to cause further ionisation. The number of excitations produced is proportional to the width of the EL region, and therefore the resulting scintillation produced by de-excitation, which we call \stwo, amplifies the ionisation signal with sub-Poissonian fluctuations, yielding excellent energy resolution.

The EL region is defined by two transparent grids separated by $\sim$10 mm, the anode itself, held at ground, and the gate, which is held at a voltage that produces a reduced electric field in this region on the order of 2 kV/cm/bar. In typical operations, 100-1000 photons are emitted per ionisation electron. Photons emitted in the direction of the anode intersect a plane, called the tracking plane, equipped with an array of light-sensitive detectors (SiPMs), separated at some pitch in the X and Y dimensions. Tracks are reconstructed combining the information of the SiPMs and the timing information provided by the TPC. This ability to reconstruct tracks is a distinctive feature of the NEXT technology, which provides a strong handle to suppress backgrounds.

\subsection{Reconstruction and background rejection}
In the standard NEXT analysis chain, a sequence of selection cuts is applied to reject background events. This consists of an initial energy cut near \Qbb, followed by a cut requiring complete containment within a specified fiducial volume. The fiducial cut removes events due to external charged particles entering the detector and gamma rays that are only partially contained. Then, a ``single-track'' cut is applied, which selects events consisting of only one continuous chain of ionization, thereby removing multiple separated energy depositions due to Compton scattering, which are typically produced by high-energy gamma ray backgrounds. This series of cuts leaves only fully-contained single-track events near \Qbb, which could still be produced by gamma rays such as those from the 2.4 MeV and 2.6 MeV lines of $^{214}$Bi and $^{208}$Tl, respectively. From here, one can further discriminate between one-electron (background) and two-electron (signal) events using the topology of the remaining single track, noting that electrons deposit a significant amount of their energy near the end of their trajectory (the Bragg peak). This forms a distinctive ``blob'' of energy at the end of each electron track, one blob in the case of one electron and two blobs in the case of two electrons. This is called the topological signature. 

The standard analysis chain in NEXT for rejecting backgrounds in a \bbonu\ search consists of a sequence of selection criteria. First, an energy cut is made selecting events near \Qbb. Then, a second cut is made which requires these events to be fully contained within a fiducial volume, away from detector surfaces, which rejects external charged particles and partially contained gamma-ray events. Third, a ``single-track cut'' is applied to select events that consist of a single, continuous ionization trail, efficiently removing multi-site Compton events. This cut is highly effective at suppressing backgrounds from high-energy gamma rays, which often produce multiple, spatially separated Compton scatters. It leaves only single electrons of energy close to \Qbb, such as those produced by the interactions of the 2.4 MeV and 2.6 MeV $\gamma$ lines emitted by $^{214}$Bi and $^{208}$Tl, respectively. It is then possible to further separate between one-electron (background) and two-electron (signal) events by taking advantage of the fact that electrons propagating in the dense gas deposit a significant amount of their energy suddenly at the end of their trajectory (the Bragg peak), forming a distinctive ``blob'' of energy, in the case of one electron, and two blobs in the case of two electrons. This is called the topological signature. 

In this paper, we study the intrinsic separation power of the topological signature and its dependence and tradeoffs with operational parameters. To achieve this, we simplify the analysis by directly comparing simulated signal events against a background sample of pure one-electron tracks of the same energy. With this approximation, we do not need to apply the preceding single-track cut, as our background sample is already idealized to contain only single-track events by construction. This approximation isolates the performance of the topological discrimination from the efficiencies of other cuts. For two key configurations, we perform the full gamma ray background generation, single-track cut, and topological classification to confirm the validity of this approximation. For the classification task, we employ an SCN, which is exceptionally well-suited for analyzing the sparse, 3D topological information of the events.

\subsection{Parameter optimisation}\label{ss.parameters}
We now consider how the operational parameters, as well as the design of the tracking plane, affect the topological signature. The two most important effects here are, 1) the impact of the diffusion of the ionisation electron trail on its way to the anode, and 2) the pitch between SiPMs which defines the granularity with which the EL signal will be sampled. Both diffusion and pitch can, in principle, be tuned in NEXT.

The diffusion in pure xenon at high pressure is large, of the order of 10 mm/$\sqrt{L}$~ in the transverse dimensions and about 3 mm/$\sqrt{L}$~ in the longitudinal dimension, where $L$ is the drift length in meters ($L = 1.187$ m in NEXT-100) \cite{Simon:2018vep}. The NEXT collaboration has studied in great detail potential ways to reduce it. One of the most promising possibilities is using a mixture of Xe-He (85/15\%)~\cite{Felkai_thesis:2022,Felkai:2017oeq,NEXT:2019oxh,McDonald_2019}, which reduces transverse diffusion to about 2 mm/$\sqrt{L}$. 

The choice of the pitch has a dramatic impact on the detector design, since the number of SiPMs and electronic channels scales with the inverse square of the pitch. Thus, there is a tension between achieving fine sampling (small pitch) and reducing costs and complexity (larger pitch). In the NEXT-White detector, the tracking plane had a diameter of about \mbox{50 cm}, and a pitch of \mbox{10 mm} resulted in 1792 SiPMs (and electronic channels). The diameter of NEXT-100 is \mbox{$\sim$ 100 cm}, and keeping the same pitch would have required 7,168 channels. Instead, the decision was made to increase the pitch to \mbox{15.5 mm} as a tradeoff for a more manageable (3584) number of sensors. The concern then is the price that must be paid for that choice, in particular with regard to the future NEXT-HD detector, with a diameter of \mbox{$\sim$ 4 m}, which would require $\sim 1.3 \times 10^5 $ channels for a pitch of \mbox{10 mm}.  

To evaluate the impact of diffusion in the performance of the topological signature we study the difference in performance between pure xenon and a Xe/He (85/15) mixture. To evaluate the impact of pitch we compare 10 and 15 mm, corresponding to the choices made for NEXT-White and NEXT-100, and we also evaluate a more costly 5 mm pitch. For these studies we use the full simulation and reconstruction of the NEXT-100 detector as input to the SCN.

A third parameter relevant for the topological signature is the operational pressure. Higher pressures result in shorter tracks, and thus a higher probability to confuse the two blobs, while lower pressures result in longer tracks and therefore more spatially distinct topological features. In this study we have also considered two scenarios, one of low pressure (\mbox{5 bar}) and another one of high pressure (\mbox{13.5 bar}) corresponding to the standard operational pressure of NEXT-100. While the higher pressure is most suitable for future tonne-scale NEXT detectors~\cite{NEXT:2020amj}, the lower pressure is also relevant, as operation at lower pressure may occur during an initial phase before the full inventory of enriched xenon is available.

This paper is organised as follows: section \ref{s:NEXT100} gives an overview of NEXT-100, the detector on which the present study is focused. Section \ref{s:montecarlo} describes the Monte Carlo simulation of the signal and background events, and section \ref{s:results} details the results of the application of the network to the various simulated configurations, with concluding remarks in section \ref{s:conclusions}. The architecture of the SCN used to perform the classification is detailed in \ref{s:network}, and the use of single-electron events as background samples in the study is discussed in \ref{s:single_electrons}.

\section{The NEXT-100 Detector}\label{s:NEXT100}
 
 \begin{figure*}[!h]
     \centering
     \includegraphics[scale=0.44]{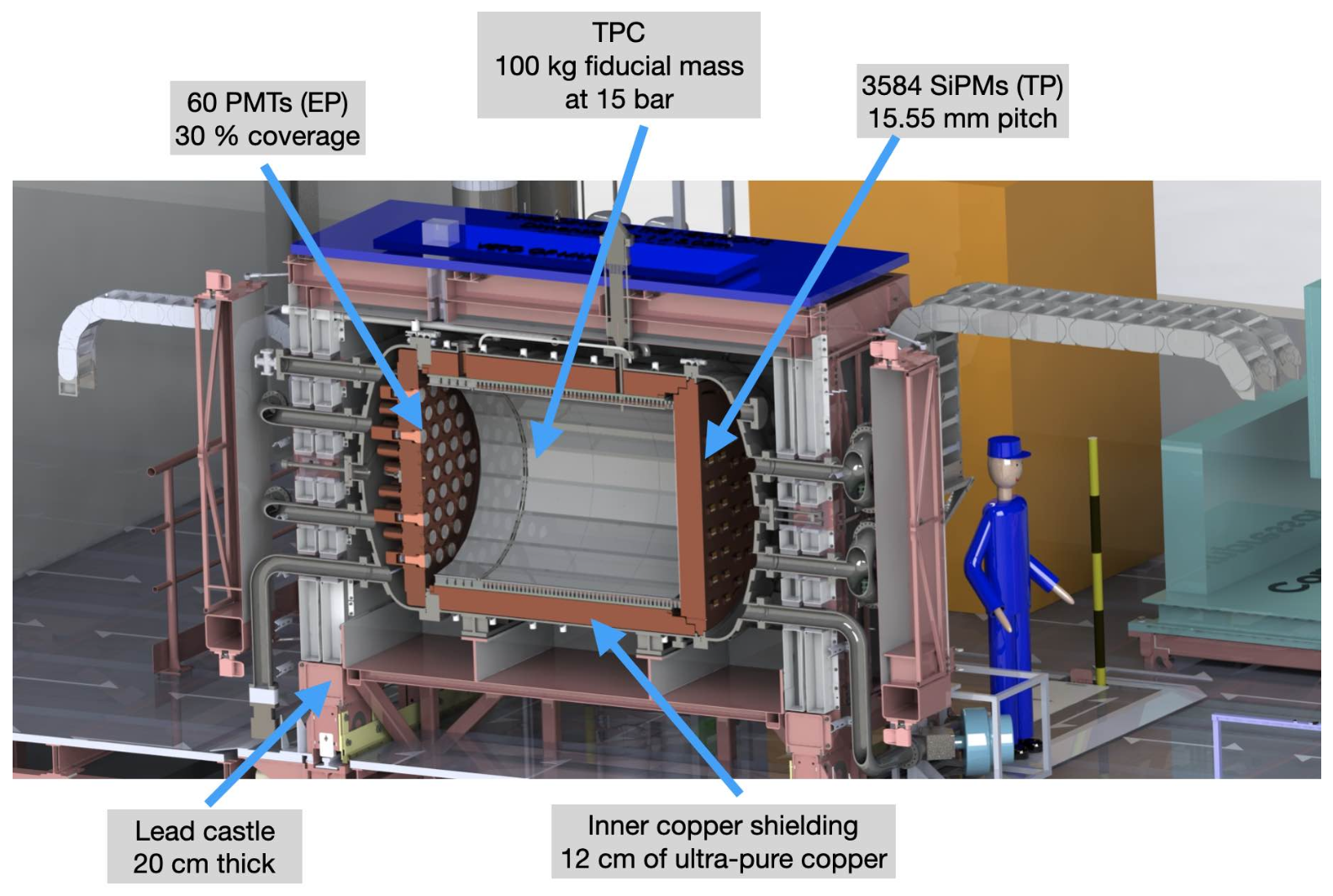}
     \includegraphics[scale=0.615]{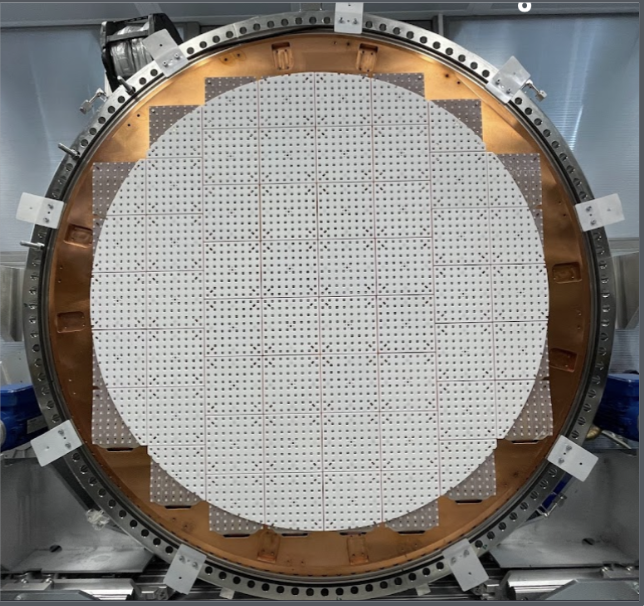}
     \includegraphics[scale=0.26]{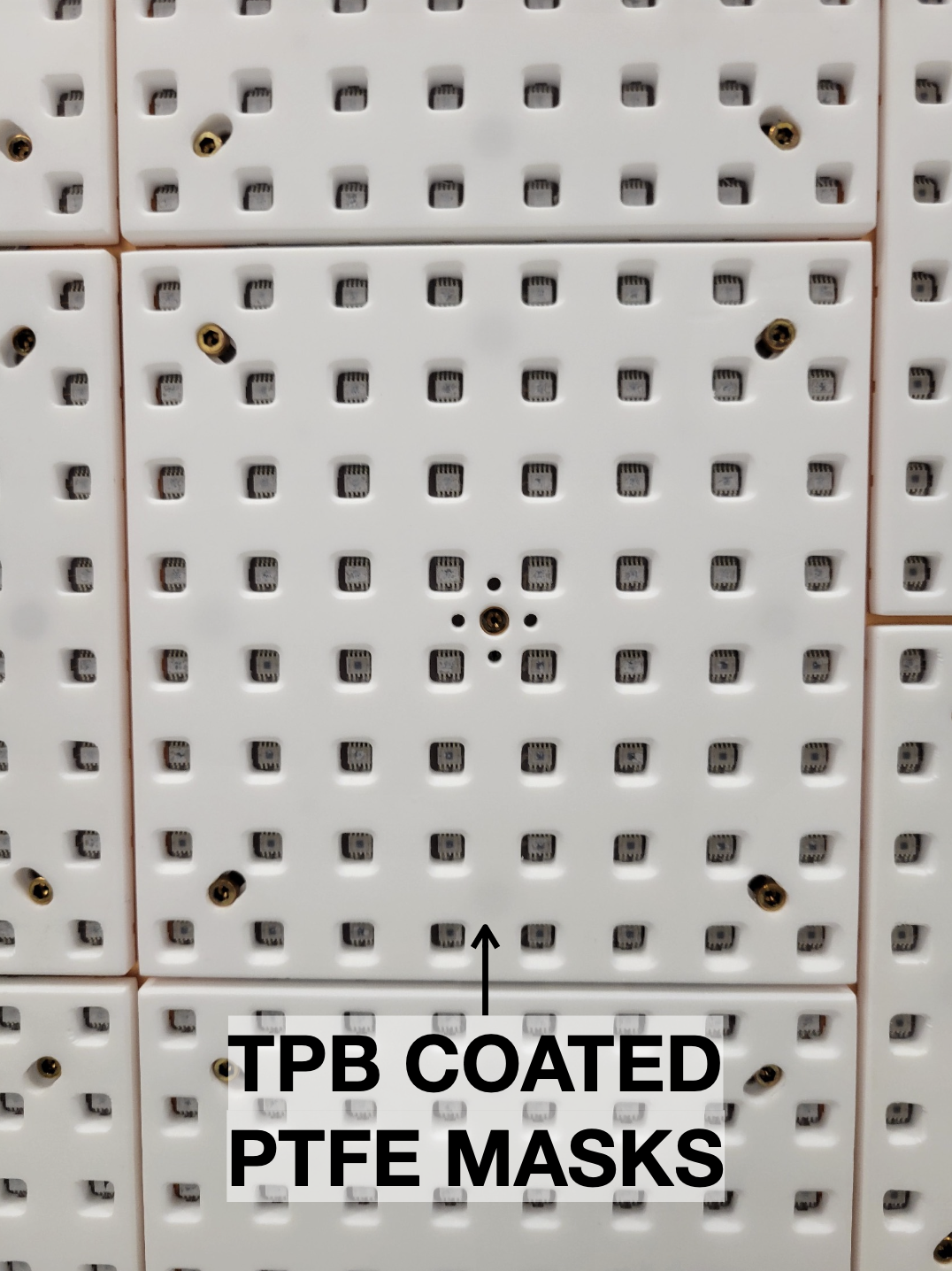}
     \caption{(Top) Schematic of the NEXT-100 detector, showing the field cage, tracking plane, energy plane (EP), and shielding layers. (Bottom) Photograph of the tracking plane (TP, left) and a close-up of a small region (right, from \cite{NEXT:2025yqw}), currently configured with 15.55 mm SiPM pitch.}
     \label{fig:NEXT100}
\end{figure*}

 The NEXT-100 detector is an asymmetric HPXe-EL TPC currently operating at the LSC~\cite{NEXT:2025yqw}. In NEXT-100, the primary scintillation (\sone) is  recorded by a plane of large photomultipliers (PMTs), located behind a transparent cathode (called the energy plane). The delayed electroluminescent signal (\stwo) is also recorded by the energy plane, as well as by the tracking plane described in Section \ref{ss.parameters}. The function of the energy plane is to record the start time of the event ($t_0$), marked by the arrival of \sone, as well as measuring the total energy deposited in the chamber, which is proportional to the integrated \stwo\ signal. The tracking plane, on the other hand, allows for the reconstruction in the XY plane of the coordinates of the event. This XY position, together with the Z coordinate (obtained from the time difference between \sone\ and \stwo), allows for the 3D reconstruction of each energy cluster deposited in the chamber with a granularity along the Z axis of the order of 1 mm. Since 2-3 MeV electrons travel significant distances in high pressure gas (e.g., around 15 cm for energies near \Qbb), the result is that an HPXe TPC can record a 3D ``picture'' of electrons travelling in the gas. To maximize the detection of the vacuum ultraviolet (VUV) light emitted by xenon, the wavelength-shifting compound tetraphenyl-butadiene (TPB) is used extensively. It absorbs the VUV light emitted by xenon and re-emits it with a peak wavelength of 420 nm, suitable for detection with high quantum efficiency by the photosensors. The SiPMs are coated with TPB, as well as the reflective PTFE panels lining the field cage and the PTFE masks placed over the regions of the tracking plane not covered by SiPMs. The PMTs are placed behind sapphire windows which are also coated with TPB. Figure \ref{fig:NEXT100} shows the major parts of the detector, including the tracking plane. A detailed description can be found in \cite{NEXT:2025yqw}.

\section{Monte Carlo Simulation}\label{s:montecarlo}
The results presented in this study are based on a detailed Monte Carlo (MC) simulation of the NEXT-100 detector using Nexus~\cite{MartinAlbo2015,NEXUS}, a Geant4-based~\cite{Agostinelli:2002hh} simulation framework developed for NEXT detectors, and Invisible Cities~\cite{IC}, a reconstruction and analysis package developed within NEXT. To simulate large numbers of events, key parts of the simulation involving propagation of optical photons are simplified in a fast simulation procedure which we call ``sensim''. The final result of the simulation is a collection of ``hits'', each containing coordinates $(x,y,z)$ and energy $E$ which are then binned into 3D ``voxels'' that can be input to the SCN for track classification.

\subsection{Monte Carlo simulation pipeline}\label{ss:simulation}
A multi-stage Monte Carlo simulation pipeline was employed, which begins with a Geant4 simulation using Nexus to generate physics events, followed by the ``sensim'' fast detector response simulation.

\subsubsection*{Geant4 detector simulation (Nexus)}
The initial stage of the simulation is Nexus, which implements a detailed geometry of the NEXT-100 detector, including the pressure vessel, internal copper shielding, field cage, and photosensor planes. Signal events were \bbonu\ decays of $^{136}$Xe, in which two electrons were emitted from a common vertex with a total kinetic energy equal to $Q_{\beta\beta} = 2.458$ MeV. The momentum of the two electrons was generated according to a distribution computed by the DECAY0 code~\cite{Ponkratenko2000}, a tool that generates kinematics for \bbonu\ decays. For the main parameter scan (section \ref{ss:param_scan}), background events were single-electrons with energy 2.458 MeV, generated uniformly within the active volume of the detector. A more detailed dataset of gamma ray events from the decays of $^{214}$Bi (2.448 MeV) and $^{208}$Tl (2.615 MeV), launched from different volumes of the detector geometry, was produced to repeat the study for two important configurations (see \ref{s:single_electrons}). The output of this initial Nexus stage is a set of ``true'' ionization hits for each event, each consisting of a deposited energy, 3D location, and time. This collection of hits serves as the ground truth for each event before detector response modeling is applied.

As simulating every optical photon produced in the initial excitation of the xenon medium and in the electroluminescence process would be computationally prohibitive, the light response is pre-computed and stored in several databases using dedicated Nexus simulations in which all optical photons were produced and tracked from all locations in the active region of the detector geometry. These databases include Light Tables and Point Spread Functions (PSFs). The light tables contain the probability of detection of S1 light produced at a location $(x,y,z)$ in the active volume, and the probability of detection of S2 light produced at a location $(x,y)$ in the EL plane for each PMT of the energy plane. The PSFs are analogous maps for the tracking plane, storing the expected charge distribution for a point-like EL light source at a location $z$ in the EL gap. Rather than storing the distribution at each point for every sensor, the PSF was computed as a function of the distance $d$ from the emission point. The fast simulation uses these databases as inputs to avoid having to model the full transport of each optical photon for every event.

\subsubsection*{Sensim: fast simulation of detector response}
The true ionization hits from Nexus were processed through a direct energy-to-sensor response simulation ``sensim''. This approach begins with true ionization hits and simulates electron drift, diffusion, and EL production. The PSFs are then used to compute the energy corresponding to the signal observed on each SiPM, bypassing the reconstruction of sensor waveforms. This produces a simulated dataset that takes into account diffusion and the spread of the produced secondary scintillation but does not require modeling of the sensor responses. The goal of this strategy was to remove all effects due to modeling of the sensors, electronics, and waveform-based reconstruction and focus on the physics affecting the measured track topologies.

\subsection{Dataset preparation}
\begin{figure}[!h]
    \centering
    \includegraphics[scale=0.5]{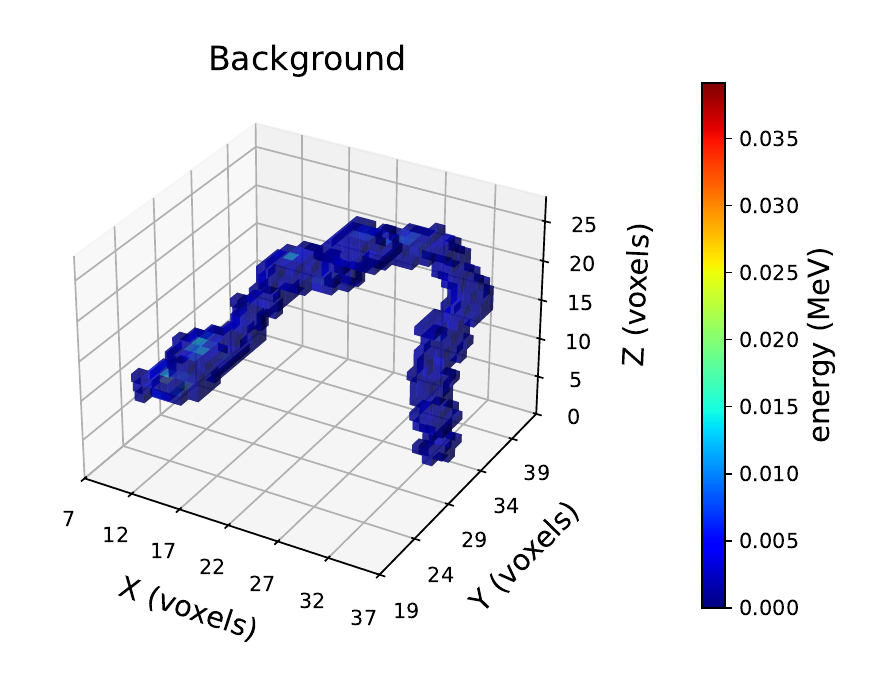}
    \includegraphics[scale=0.5]{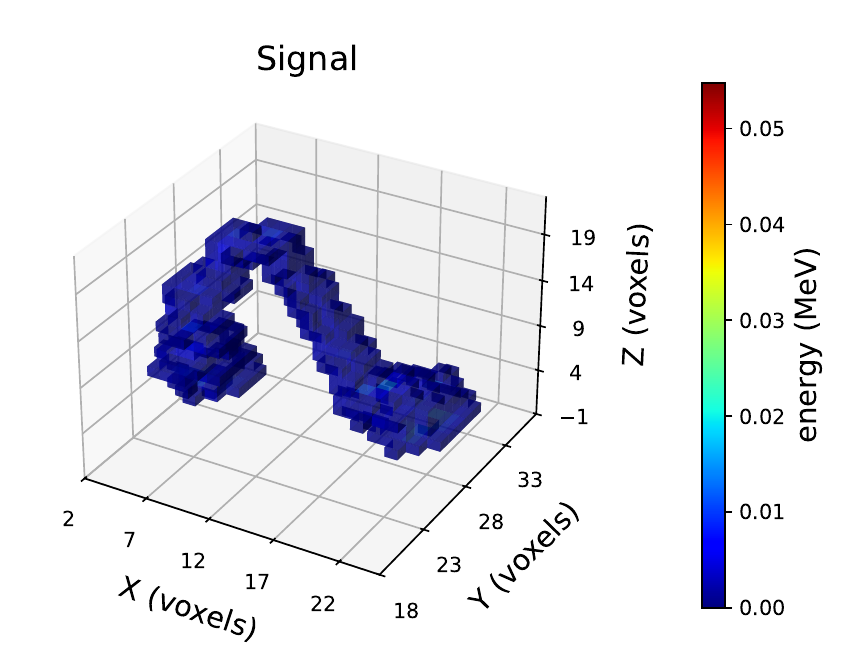}
    \caption{A voxelised simulated single-electron background event (top) and a two-electron $0\nu\beta\beta$ event (bottom).}
    \label{fig:mc_examples}
\end{figure}

To generate the final inputs to the SCN, the hits from each event were voxelised into a 3D grid as shown in Fig. \ref{fig:mc_examples}, and the energy deposited in each voxel was summed. The voxel size in the transverse (XY) plane was chosen to match the SiPM pitch being studied, while the size in the drift (Z) direction was chosen to be approximately 10 mm (5 mm in the case of 5 mm pitch). To ensure the network learns topological features irrespective of the event's total energy, the energy in all voxels for a given event was normalized to 1.

The detector was configured with several variations in geometry and gas media. In particular, the tracking plane was modeled with SiPM pitches of 5 mm, 10 mm and 15 mm. This allowed for a direct study of the trade-off between spatial resolution and construction complexity. Studies were performed at 5 bar and 13.5 bar xenon gas pressures. To explore the effects of reduced diffusion, the same studies were performed at all pitches and pressures using a gas mixture of 85\% xenon and 15\% helium. A useful reference point was also evaluated at each gas pressure by studying the true Monte Carlo tracks voxelised with a size of 2 mm, corresponding to the case of no diffusion and near-perfect reconstruction.

All events were generated uniformly within the active volume, and a final cut on the summed energy of all voxels was made, requiring $E \in [2.4, 2.55]$ MeV spanning a range that includes $Q_{\beta\beta}$. A total of approximately 100k signal and 100k background events were simulated for each configuration, with the datasets split into training (70\%), validation (20\%), and test (10\%) subsets. The SCN was trained from scratch for each dataset using Pytorch~\cite{Pytorch}, using the training subset in the network optimisation, the validation subset to select the optimal training epoch, and the test set for the final evaluation (see Fig. \ref{fig:loss_curve_and_ROC}, top, and \ref{ss:network_training}). The study was performed for a total of 16 datasets, including: 12 main configurations (2 pressures $\times$ 2 gas mixtures $\times$ 3 SiPM pitches) + 2 reference studies using Monte Carlo truth at 2 mm voxel size, and 2 key configurations using radiogenic gamma events rather than single-electron events as the background sample.
\section{Results}\label{s:results}
\subsection{Parameter scan}\label{ss:param_scan}
\begin{figure}[!h]
    \centering
    \includegraphics[scale=0.5]{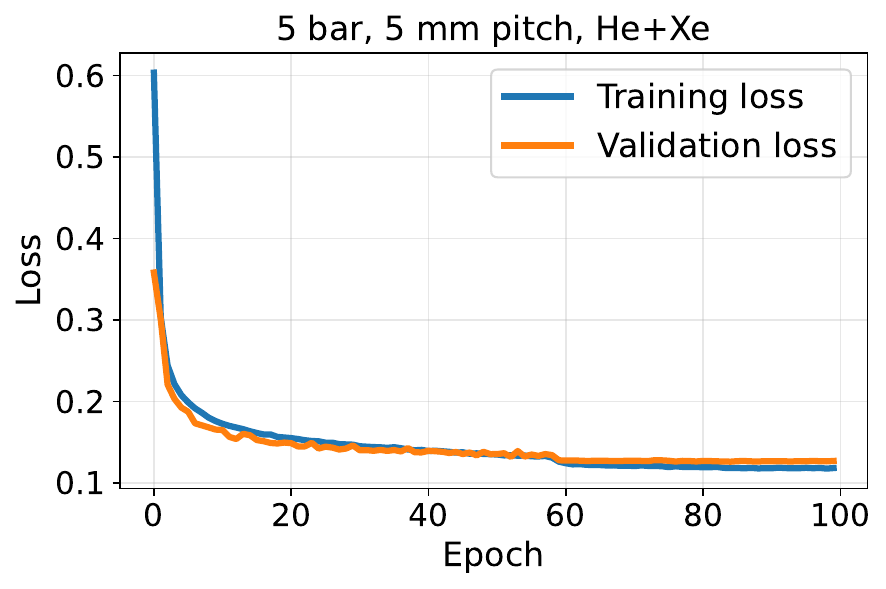}
    \includegraphics[scale=0.5]{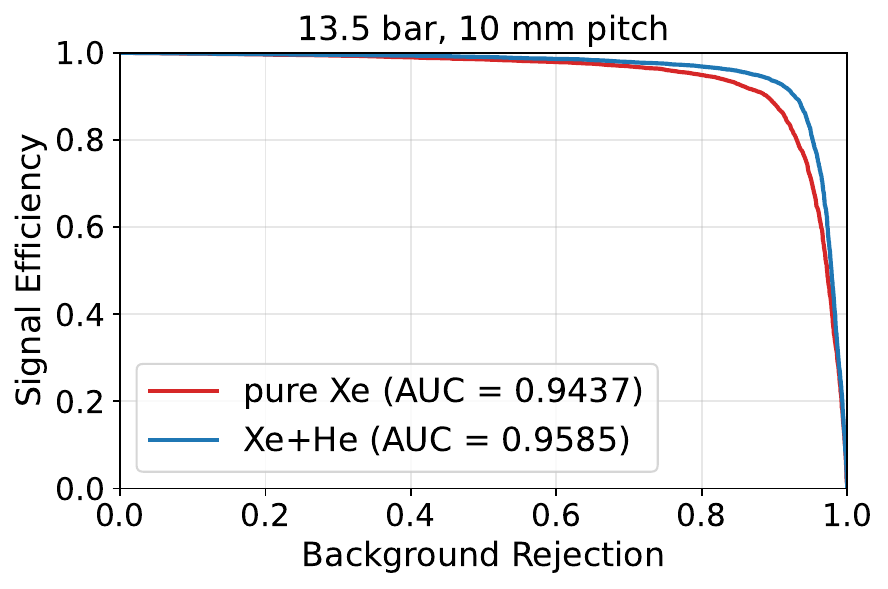}
    \caption{(Top) Example training and validation loss curves, He+Xe mixture, 5 mm pitch, 5 bar pressure. (Bottom) ROC curves for 13.5 bar, 10 mm pitch with and without He admixture.}
    \label{fig:loss_curve_and_ROC}
\end{figure}

The performance of the SCN was evaluated using two key metrics: the area under the Receiver Operating Characteristic (ROC) curve, known as the AUC, and background rejection at 80\% signal efficiency. These metrics were analyzed for datasets simulated at \mbox{5 bar} and \mbox{13.5} bar pressures, the various tracking plane pitches (\mbox{5 mm}, \mbox{10 mm}, and \mbox{15 mm}) and gas mixtures (pure xenon and helium-xenon). Some example ROC curves are shown in Fig. \ref{fig:loss_curve_and_ROC} (bottom), and a detailed breakdown of the AUC values obtained for the different configurations studied can be found in \ref{ss:AUC}.

\begin{figure*}[!h]
    \centering
    \includegraphics[scale=0.7]{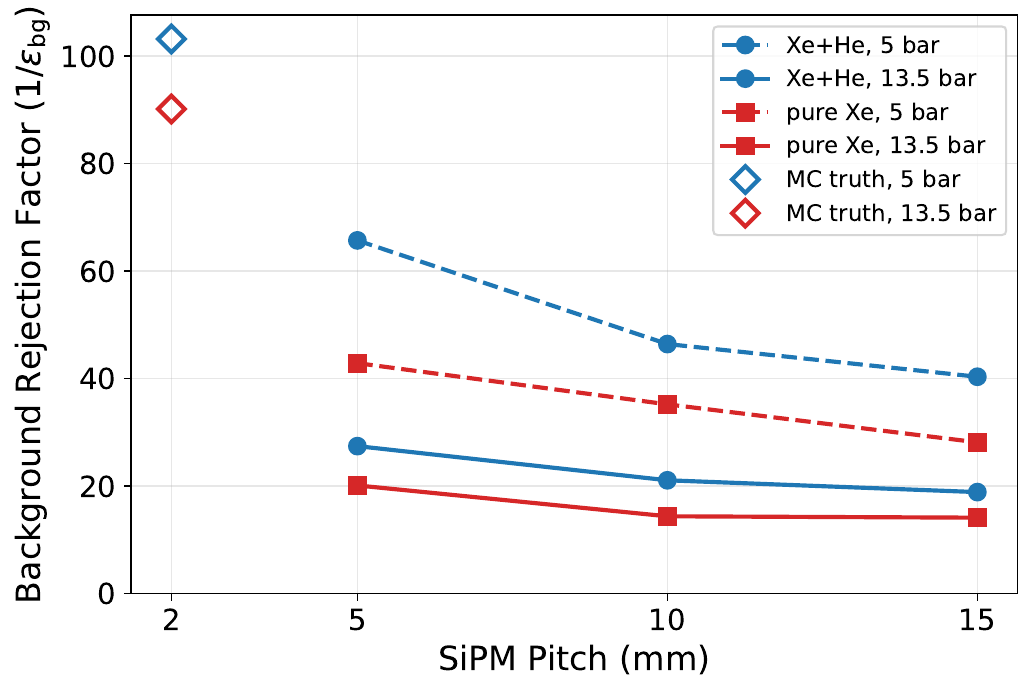}
    \caption{Background rejection factor ($1/\epsilon_{\mathrm{bg}}$) at 80\% signal efficiency as a function of SiPM pitch, for the different configurations of gas mixture (pure Xe in red, Xe+He in blue) and pressure (13.5 bar solid, 5 bar dashed). The diamond-shaped points show, for reference, the near-perfect reconstruction case of MC true hits binned directly into 2$\times$2$\times$2 mm$^3$ voxels (plotted at a pitch of 2 mm). The improvements in background rejection with each decrease in pitch are evident, as well as the improvements at lower pressure and lower diffusion (Xe+He).}
    \label{fig:rejection_vs_pitch}
\end{figure*}

Figure~\ref{fig:rejection_vs_pitch} shows the inverse of the background acceptance fraction ($1 / \epsilon_{\mathrm{bg}}$) evaluated at 80\% signal efficiency for the different configurations. Higher values indicate better rejection. These results quantify the expected trends, showing how reduced pitch, reduced diffusion (addition of He), and lower pressure tend towards better performance.

\subsection{Background rejection in key configurations}\label{ss:key_configurations}
We now attempt to put these results in the context of the total expected background rate for NEXT-like detectors. In \cite{NEXT:2020amj}, a sensitivity study was presented for a tonne-scale NEXT experiment, in which the most prominent backgrounds were identified as radiogenic gamma rays from $^{208}$Tl and $^{214}$Bi present in the detector materials. Other important backgrounds come from natural radioactivity, including decay products of $^{222}$Rn and cosmogenic $^{137}$Xe produced by neutron interactions with $^{136}$Xe. The study simulated the various background events and \bbonu\ events, along with the full analysis chain, including a wide energy cut, fiducial volume cut, single-track cut, a classical topological cut (based on identification of the two energetic ``blobs'' at the ends of the tracks), and a final region-of-interest (ROI) energy window. The simulation was done at 15 bar gas pressure, and the physics of the detection of electroluminescent photons was not simulated directly. Rather the ionization depositions produced in Geant4 were voxelised directly with voxel size 3 $\times$ 3 $\times$ 3 mm$^3$ (similar to the ``MC truth'' result shown in Fig. \ref{fig:rejection_vs_pitch}). The conclusion was an estimated background rate of 0.061 events / (tonne-year-ROI) for a 24.6\% signal efficiency. 

We consider the results of this tonne-scale sensitivity study, substituting the topological cut with our present results for two key configurations:
\begin{enumerate}
    \item[1.] Standard NEXT-100: 15 mm pitch and pure Xe. This reflects the initial operating conditions of NEXT-100.
    \item[2.] NEXT-100 upgrade: 10 mm pitch and an 85/15 Xe/He mixture. This is consistent with a future planned upgrade to NEXT-100.
\end{enumerate} 

Replacing the topological cut employed in the tonne-scale study with the values obtained in Fig. \ref{fig:rejection_vs_pitch} would result in an estimated background rate of 0.087 events/(tonne-year-ROI) for a standard NEXT-100 configuration and 0.058 events/(tonne-year-ROI) for the NEXT-100 upgrade configuration with Xe/He, in both cases with an overall signal efficiency of 26.9\%.
\section{Discussion and Conclusions}\label{s:conclusions}
The present study has employed a Sparse Convolutional Neural Network to study the optimal event classification capabilities of NEXT-100 under several different tracking plane configurations and gas mixtures. From the results obtained, we arrive at the conclusions discussed below.

First, topological rejection power improves with lower pressure. If we compare the two best cases at 5 and 13.5 bar (corresponding to the smallest pitch and the smallest diffusion), we find that the fraction of background events (at 80\% signal efficiency) in the case of 5 bar is 1.5\%, close to the intrinsic level, and 3.7\% at 13.5 bar. Thus, the lower pressure yields more than a factor of 2 less background. Note, however, that this does not translate directly to a strong statement on overall sensitivity to \bbonu, as a proper assessment would require accounting for detector size, containment efficiency, intrinsic backgrounds, and other factors which we do not address in the present study.

Second, the addition of helium to the xenon gas mixture improves background rejection considerably for a given signal efficiency, suggesting that the inclusion of helium would be advantageous if associated operational challenges, such as increased complexity in gas handling, are not prohibitive. We also note that the addition of helium would reduce the total amount of $^{136}$Xe present in the active volume, which must be taken into account when evaluating overall sensitivity.

Finally, the comparison between tracking plane pitches highlights the incremental advantage of a lower pitch. If we compare the three pitches (at 13.5 bar with HeXe, which we have identified as the best case), we find that the 5 mm pitch yields a background fraction of 3.7\%, 10 mm pitch yields 4.7\%, and 15 mm pitch yields 5.3\%. The difference between the best and worst case is less than a factor of 1.5, while the cost of the tracking plane would roughly increase by a factor $3^2 \sim 10$. Here, most likely, the balance between cost and performance would argue for a pitch of 10 mm.  

In conclusion, our study suggests that operation at high pressure with an 85/15 Xe/He mixture and a pitch of 10 mm are potentially optimal parameters for topological classification in future NEXT detectors. Details of detector geometry, radiopurity, and total mass of $^{136}$Xe will affect the ultimate sensitivity of the experiment. Improvements from 15 mm $\rightarrow$ 10 mm pitch are incremental, as are improvements from 10 mm $\rightarrow$ 5 mm. However, the cost to move from 10 mm $\rightarrow$ 5 mm is significantly higher. As 10 mm has already been demonstrated in a previous NEXT detector, it is a solid target for a future-generation experiment.\\

\appendix
\section{Network Architecture, Training, and Evaluation}\label{s:network}
The SCN is designed to efficiently process sparse voxelised data, considering only voxels with nonzero energy deposition. Therefore the amount of memory and computation required does not grow with the size of the entire sensitive volume but rather only with the size of the tracks, making the method scalable to larger detector sizes. The SCN architecture supports operations such as submanifold convolutions, which preserve sparsity by maintaining the same number of non-zero voxels, and standard sparse convolutions, which are used for downsampling and feature extraction.

\subsection{Network Configuration}
\begin{figure*}[!h]
    \centering
    \includegraphics[scale=0.3]{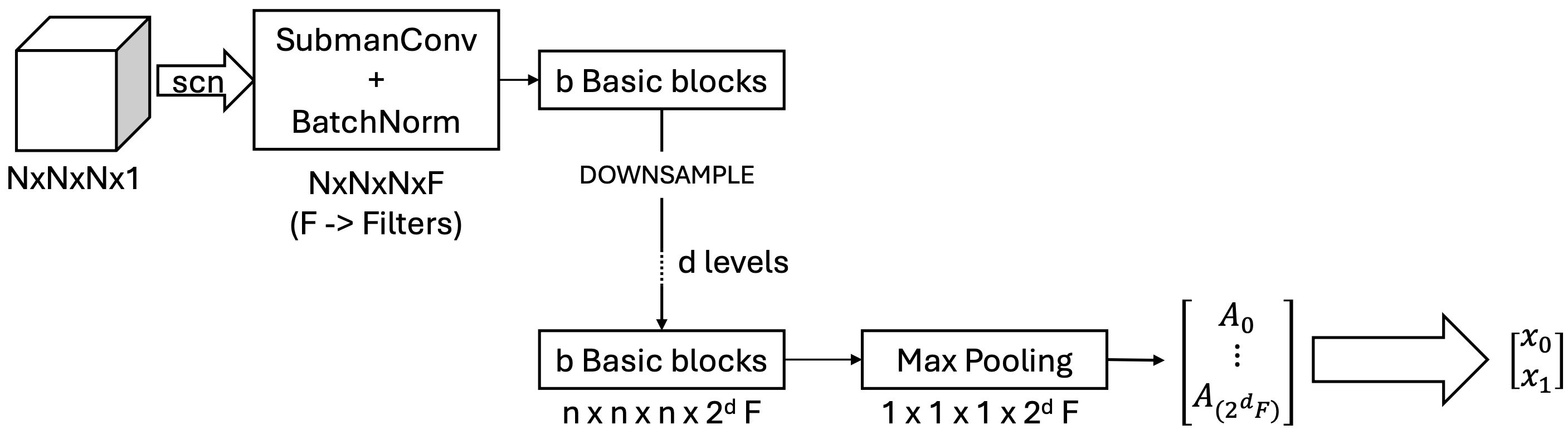}
    \vspace{0.5cm}
    \hrule
    \vspace{0.5cm}
    \includegraphics[scale=0.3]{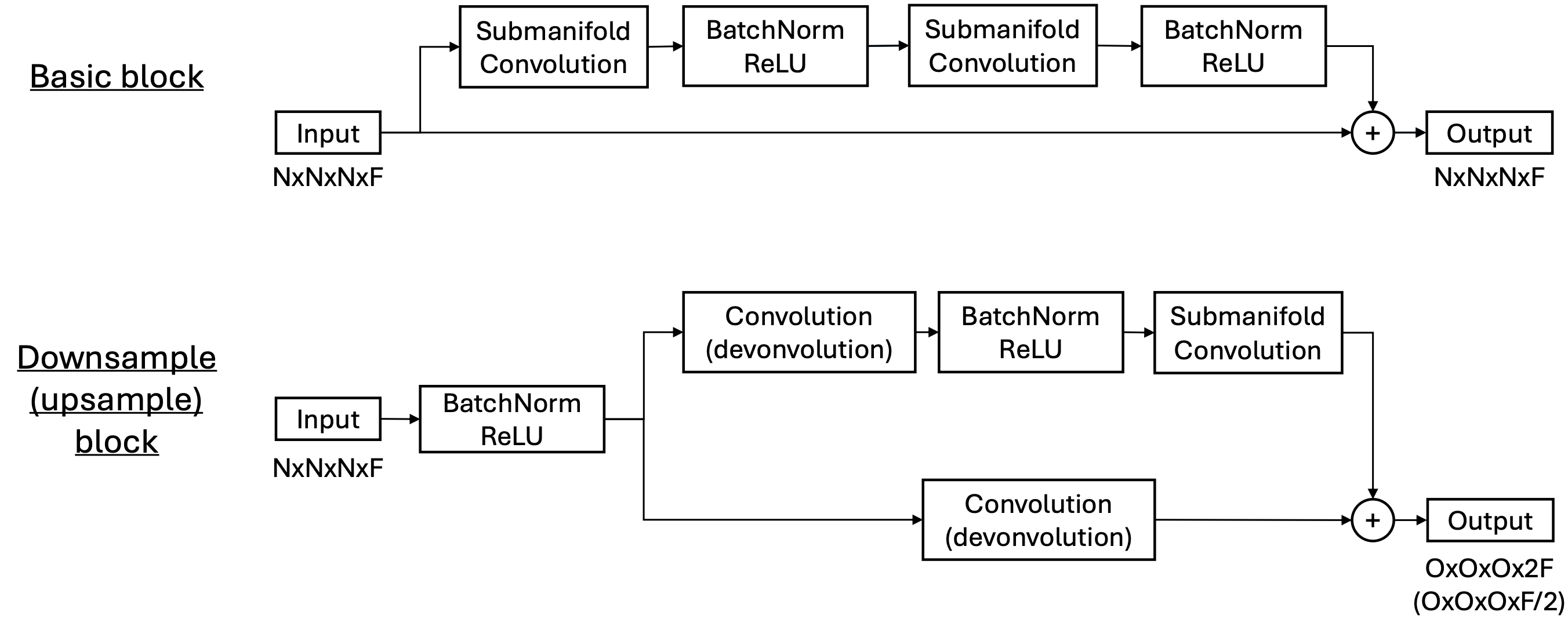}
    \caption{SCN ResNet architecture. (Top) The full SCN used in this study. (Bottom) The structure of the individual basic and downsampling blocks.}
    \label{fig:network_architecture}
\end{figure*}

The network architecture used in this study is a ResNet~\cite{He2015} built with sparse convolutional layers. The architecture is illustrated in Figure~\ref{fig:network_architecture}. The input spatial size was selected to allow for voxel coordinates inside the entire NEXT-100 volume (1 meter $\times$ 1 meter in X-Y and 1.3 meters in Z). As only certain spatial sizes were permitted by the architecture of the network, we used input dimensions of (223, 223, 223), the minimum size permitted, in most cases, but needed to increase the size to (223, 223, 287) for 5 mm voxels and (671, 671, 671) for the 2 mm voxels size employed in the Monte Carlo truth analysis. Batch normalization layers were used throughout the network, with a momentum parameter of 0.7. The structure of the network consisted of (see Fig. \ref{fig:network_architecture}):

\begin{itemize}
    \item \textbf{Initial convolution:} The first layer applies a convolution with 8 planes and a kernel size of 7 voxels.
    \item \textbf{Basic blocks:} The network consists of $d=5$ layers organized into basic residual blocks + downsampling (see below) blocks. Each group of $b=2$ blocks uses kernel sizes of $[7, 7, 5, 3, 3]$. A final group of $b=2$ basic blocks is applied at the end with kernel size 3.
    \item \textbf{Downsampling blocks:} Downsampling is applied after each of the $d=5$ groups of basic blocks, in which the spatial dimensions are reduced and the feature map dimensions duplicated at each step. Convolutions in the five downsampling layers use kernel sizes of $[7, 7, 5, 3, 3]$ and strides of $[4, 2, 2, 2, 2]$.
    \item \textbf{Fully connected layers:} After the final convolutional block, the network includes two fully connected layers with $n_{\text{linear}} = 32$ and $n_{\text{classes}} = 2$ (signal and background).
\end{itemize}

\subsection{Network training}
\label{ss:network_training}
Training and inference were performed on various Nvidia GPUs, including GeForce GTX 1080 Ti, TITAN X (Pascal), A100, V100, and RTX PRO 6000 Blackwell models. Training was continued until the loss converged, as determined qualitatively based on minimal changes in loss between epochs. An example loss curve is shown in Fig. \ref{fig:loss_curve_and_ROC} (top). The optimiser used was Adam~\cite{Kingma_2017} with a learning rate of \(10^{-3}\). To improve the network's robustness, data augmentation was applied to the voxelised events. Augmentation operations included random mirroring of events along the X, Y, and Z axes and rotation of the events about the X, Y, and Z axes.

\subsection{Area Under the Receiver Operating Characteristic Curve (AUC) Evaluation}\label{ss:AUC}
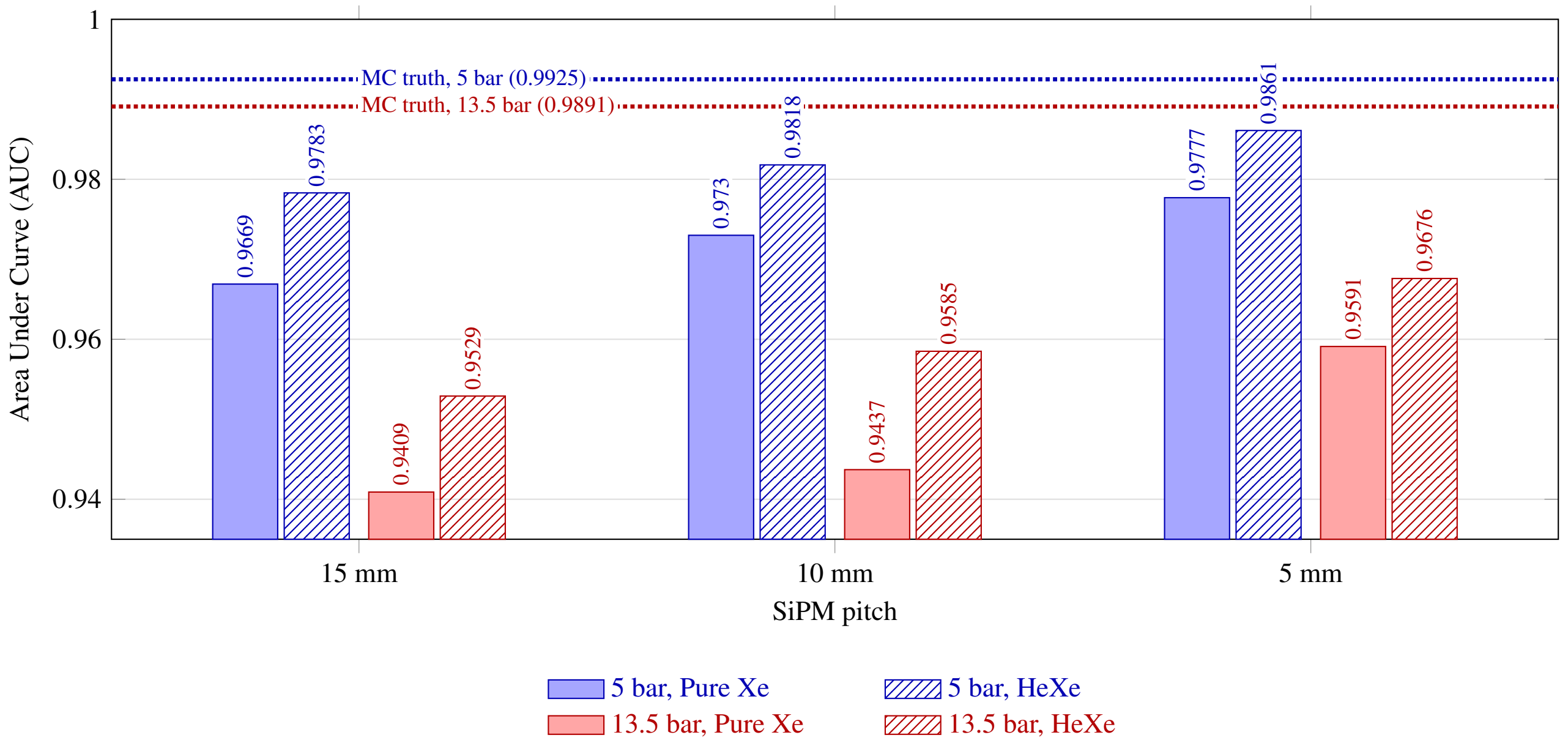
\begin{figure*}[!htbp]
    \centering

    \begin{tikzpicture}
        \begin{axis}[
            ybar,
            width=0.99\textwidth,
            height=7.2cm,
            bar width=20pt,
            ylabel={Area Under Curve (AUC)},
            xlabel={SiPM pitch},
            symbolic x coords={15 mm, 10 mm, 5 mm},
            xtick=data,
            ymin=0.935, ymax=1.000,
            ytick distance=0.02,
            ymajorgrids=true,
            grid style={draw=gray!25},
            enlarge x limits=0.26,
            tick label style={font=\small},
            label style={font=\small},
            area legend,
            legend style={
                at={(0.5,-0.24)},
                anchor=north,
                legend columns=2,
                draw=none,
                font=\small,
                /tikz/every even column/.append style={column sep=0.8cm}
            },
            legend cell align=left,
            nodes near coords,
            every node near coord/.append style={
                font=\scriptsize,
                rotate=90,
                anchor=west,
                /pgf/number format/fixed,
                /pgf/number format/precision=4,
                fill=white,
                fill opacity=1,
                text opacity=1,
                inner sep=1pt,
                outer sep=1pt,
                draw=none
            },
            clip=false
        ]

        % -------------------------
        % 5 bar, Pure Xe (blue solid)
        % -------------------------
        \addplot+[
            draw=blue!70!black,
            fill=blue!35,
            bar shift=-35pt,
            every node near coord/.append style={text=blue!70!black}
        ] coordinates {
            (15 mm, 0.9669)
            (10 mm, 0.9730)
            (5 mm, 0.9777)
        };

        % -------------------------
        % 5 bar, HeXe (blue hatched)
        % -------------------------
        \addplot+[
            draw=blue!70!black,
            fill=blue!20,
            pattern=north east lines,
            pattern color=blue!70!black,
            bar shift=-13pt,
            every node near coord/.append style={text=blue!70!black}
        ] coordinates {
            (15 mm, 0.9783)
            (10 mm, 0.9818)
            (5 mm, 0.9861)
        };

        % -------------------------
        % 13.5 bar, Pure Xe (red solid)
        % -------------------------
        \addplot+[
            draw=red!70!black,
            fill=red!35,
            bar shift=13pt,
            every node near coord/.append style={text=red!70!black}
        ] coordinates {
            (15 mm, 0.9409)
            (10 mm, 0.9437)
            (5 mm, 0.9591)
        };

        % -------------------------
        % 13.5 bar, HeXe (red hatched)
        % -------------------------
        \addplot+[
            draw=red!70!black,
            fill=red!20,
            pattern=north east lines,
            pattern color=red!70!black,
            bar shift=35pt,
            every node near coord/.append style={text=red!70!black}
        ] coordinates {
            (15 mm, 0.9529)
            (10 mm, 0.9585)
            (5 mm, 0.9676)
        };

        \legend{
            \textcolor{blue!70!black}{5 bar, Pure Xe},
            \textcolor{blue!70!black}{5 bar, HeXe},
            \textcolor{red!70!black}{13.5 bar, Pure Xe},
            \textcolor{red!70!black}{13.5 bar, HeXe}
        }

        % -------------------------
        % MC-truth lines across full axis width
        % -------------------------
        % y = 0.9925 corresponds to rel y = (0.9925-0.935)/(1.00-0.935)=0.88462
        \draw[blue!70!black, line width=0.45mm, densely dotted]
            (rel axis cs:0,0.88462) -- (rel axis cs:1,0.88462);

        \node[
            text=blue!70!black,
            font=\scriptsize,
            fill=white,
            inner sep=1pt,
            anchor=west
        ] at (rel axis cs:0.17,0.88462)
            {MC truth, 5 bar (0.9925)};

        % y = 0.9891 corresponds to rel y = (0.9891-0.935)/(1.00-0.935)=0.83231
        \draw[red!70!black, line width=0.45mm, densely dotted]
            (rel axis cs:0,0.83231) -- (rel axis cs:1,0.83231);

        \node[
            text=red!70!black,
            font=\scriptsize,
            fill=white,
            inner sep=1pt,
            anchor=west
        ] at (rel axis cs:0.17,0.83231)
            {MC truth, 13.5 bar (0.9891)};

        \end{axis}
    \end{tikzpicture}

    \caption{
        AUC values for different SiPM pitch configurations. Blue bars correspond to 5 bar and red bars to 13.5 bar. Within each pressure, solid bars denote pure Xe and hatched bars denote HeXe. The horizontal dotted lines show the results corresponding to MC truth-based voxelisation at 2$\times$2$\times$2 mm$^{3}$.
    }
    \label{fig:auc_barchart}
\end{figure*}

The AUC gives the probability that the network scores a signal event as more ``signal-like'' than a background event, and therefore it provides a single-value metric for evaluating the network's ability as a classifier. An AUC of 1.0 represents a perfect classifier, while random guessing would yield an AUC of 0.5. Figure~\ref{fig:auc_barchart} summarizes the AUC values obtained for the different detector configurations at 5 bar and 13.5 bar. The AUC for an analysis of the Monte Carlo true energy hits voxelised to 2$\times$2$\times$2 mm$^3$ is also shown for reference. As expected, performance improves significantly with the addition of He gas due to the reduced electron diffusion. The overall performance is notably better at 5~bar than at 13.5~bar, with AUCs at 5 bar 1-3 percentage points higher for a given gas mixture and pitch. This is due to the more compact nature of particle tracks at increased pressure, making the features of the tracks more difficult to distinguish and the smearing due to diffusion more critical.

\section{Single-electron tracks as a proxy for radiogenic backgrounds}
\label{s:single_electrons}
A significant concern is whether the use of single-electron tracks as ``background'' events, which more closely resemble the energetic electrons emitted in cosmogenic-induced $^{137}$Xe decays, is representative of the results obtained for background tracks produced by high-energy gamma rays. The topology of a simulated energetic electron may not match exactly that of the more complex topologies produced by high-energy gamma rays, which can include Compton scattering, pair production, and photoelectric interactions. While multi-track events can be discarded using a well-designed single-track cut, in practice this cut will not always select topologies that correspond exactly to tracks produced by single-electron events. The motivation for using single electrons as background events in this study is that the simulation and interpretation of the results are significantly more straightforward.

Therefore, to verify that this is a good approximation to single-track events selected from realistic backgrounds, we have produced a dataset of $^{214}$Bi and $^{208}$Tl events at 13.5 bar and, from these events, selected topologies passing a single-track cut implemented using the true Monte Carlo hits. The cut eliminated approximately 80\% of background events. We then simulated the sensor responses using ``sensim'' for the two key configurations discussed in section \ref{ss:key_configurations}: pure Xe, 15 mm pitch (standard NEXT-100) and Xe/He, 10 mm pitch (NEXT-100 upgrade). For these two datasets, both the signal and background samples were
restricted to an energy window of $E \in [2.3, 2.55]$ MeV, rather than the
$[2.4, 2.55]$ MeV window used for the single-electron datasets, as the majority of the events in the background generation had energies lower than 2.4 MeV. Since the total voxel energy of each event is
normalized before entering the network, the absolute energy scale does not enter directly into the classification.

\begin{figure}[!h]
    \centering
    \includegraphics[scale=0.55]{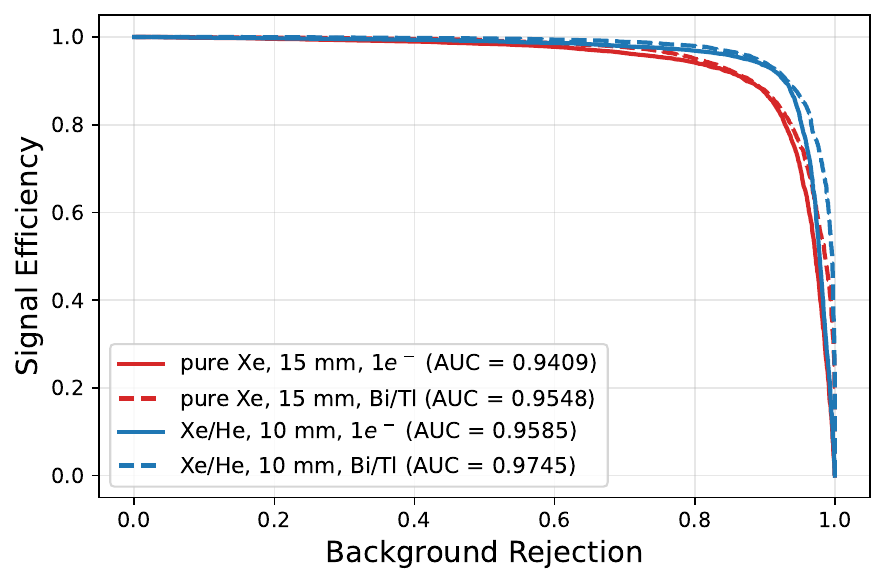}
    \caption{Comparison of results for single-electron and realistic ($^{214}$Bi and $^{208}$Tl) gamma ray backgrounds. ROC curves for each background type are shown for 13.5 bar pure Xe at 15 mm pitch and for HeXe at 10 mm pitch.}
    \label{fig:compare_single_electron}
\end{figure}

We trained an SCN in each case using these events as the background sample in place of our single-electron background sample. As shown in Figure~\ref{fig:compare_single_electron}, the results for both background event types are similar, validating our single-electron approximation. Table~\ref{tab:single_e_vs_tlbi} summarizes the background rejection at 80\% signal efficiency for both configurations and background event types.

\begin{table}[!h]
    \centering
    \begin{tabular}{lcc}
        \hline
         & Pure Xe, 15 mm & Xe/He, 10 mm \\
        \hline
        Single-electron & 7.1\% (14.1) & 4.7\% (21.1) \\
        $^{208}$Tl + $^{214}$Bi & 6.2\% (16.1) & 3.4\% (29.8) \\
        \hline
    \end{tabular}
    \caption{Background acceptance $\epsilon_{\mathrm{bg}}$ at 80\% signal
    efficiency (with the corresponding rejection factor
    $1/\epsilon_{\mathrm{bg}}$ in parentheses) for the two key
    configurations, comparing the single-electron background sample with
    the realistic radiogenic ($^{208}$Tl + $^{214}$Bi) background sample.}
    \label{tab:single_e_vs_tlbi}
\end{table}

\noindent\textbf{Funding}\;The NEXT Collaboration acknowledges support from the following agencies and institutions: the European Research Council (ERC) under Grant Agreements No. 951281-BOLD and 101039048-GanESS; the European Union’s Framework Programme for Research and Innovation Horizon 2020 (2014–2020) under Grant Agreement No. 860881-HIDDeN; the MCIN/AEI of Spain and ERDF A way of making Europe under grants PID2021-125475NB and RTI2018-095979, and the Severo Ochoa and Mar\'ia de Maeztu Program grants CEX2023-001292-S, CEX2023-001318-M and CEX2018-000867-S; the Generalitat Valenciana of Spain under grants PROMETEO/2021/087, ASFAE/2022/028, ASFAE/2022/029, CISEJI/2023/27 and CIDEXG/2023/16; the Department of Education of the Basque Government of Spain under the predoctoral training program non-doctoral research personnel; the Spanish la Caixa Foundation (ID 100010434) under fellowship code LCF/BQ/PI22/11910019; the Portuguese FCT under project UID/FIS/04559/2020 to fund the activities of LIBPhys-UC; the Israel Science Foundation (ISF) under grant 1223/21; the Pazy Foundation (Israel) under grants 310/22, 315/19 and 465; the US Department of Energy under contracts number DE-AC02-06CH11357 (Argonne National Laboratory), DE-AC02-07CH11359 (Fermi National Accelerator Laboratory), DE-FG02-13ER42020 (Texas A\&M), DE-SC0019054 (Texas Arlington) and DE-SC0019223 (Texas Arlington); the US National Science Foundation under award number NSF CHE 2004111; the Robert A Welch Foundation under award number Y-2031-20200401.\\

\noindent\textbf{Acknowledgements}\;We are grateful to the Laboratorio Subterr\'aneo de Canfranc for hosting and supporting the NEXT experiment. AI tools were used in this study to assist in the writing of computing code. Results were reviewed by the authors who assume full responsibility for the results presented here. The paper was fully written by the authors with no AI intervention.\\

\noindent\textbf{Data Availability Statement}\;The datasets generated and/or analysed during the current study are available from the corresponding author on reasonable request.\\

\noindent\textbf{Code Availability Statement}\;The code/software generated and/or analysed during the current study is available from the corresponding author on reasonable request.

\bibliographystyle{spphys}
\bibliography{NextRefs}

\end{document}